\documentclass[preprints,article,accept,moreauthors,pdftex]{mdpi} 

\firstpage{1} 
\pubvolume{xx}
\issuenum{1}
\articlenumber{5}
\pubyear{2020}
\copyrightyear{2020}
\history{} 
\usepackage{bookmark}
\usepackage{amsmath} 
\usepackage{amssymb}

\Title{Statistical estimates of the pulsar glitch activity}

\Author{Alessandro Montoli $^{1,2}$\orcidA{}*, Marco Antonelli $^{3}$\orcidB{}*, Brynmor Haskell $^{3}$\orcidC{}  and Pierre Pizzochero $^{1,2}$}

\AuthorNames{Alessandro Montoli, Marco Antonelli, Brynmor Haskell and Pierre Pizzochero}

\address{%
$^{1}$ \quad Dipartimento di Fisica, Universit\`a degli Studi di Milano, Via Celoria 16, 20133 Milano, Italy\\
$^{2}$ \quad Istituto Nazionale di Fisica Nucleare, sezione di Milano, Via Celoria 16, 20133 Milano, Italy\\
$^{3}$ \quad Nicolaus Copernicus Astronomical Center of the Polish Academy of Sciences, ul. Bartycka 18, 00-716 Warszawa, Poland
}

\corres{Email: \url{alessandro.montoli@unimi.it} (A.M.), \url{mantonelli@camk.edu.pl}  (M.A.)}

\abstract{
A common way to calculate the glitch activity of a pulsar is an ordinary linear regression of the observed cumulative glitch history. This method however is likely to underestimate the errors on the activity, as it implicitly assumes a (long-term) linear dependence between glitch sizes and waiting times, as well as equal variance, i.e., homoscedasticity, in the fit residuals, both assumptions that are not well justified from pulsar data.
In this paper, we review the extrapolation of the glitch activity parameter and explore two alternatives: the relaxation of the homoscedasticity hypothesis in the linear fit and the use of the bootstrap technique.
We find a larger uncertainty in the activity with respect to that obtained by ordinary linear regression, especially for those objects in which it can be significantly affected by a single glitch.
We discuss how this affects the theoretical upper bound on the moment of inertia associated with the region of a neutron star containing the superfluid reservoir of angular momentum released in a stationary sequence of glitches.
We find that this upper bound is less tight if one considers the uncertainty on the activity estimated with the bootstrap method and allows for models in which the superfluid reservoir is entirely in the crust. 
}

\begin{document}
%%%%%%%%%%%%%%%%%%%%%%%%%%%%%%%%%%%%%%%%%%

\section{Introduction}

To date, pulsar glitches are considered the most striking macroscopic manifestation of the presence of a neutron superfluid in the inner crust and outer core of neutron stars (see, e.g., \cite{haskell_melatos2015} for a recent review). 
According to the current understanding, a rotating neutron star is comprised of two components, a normal one---strongly coupled to the magnetic field of the star and observed from Earth---and a superfluid one---which lags behind the normal one during the spin-down process \cite{baym+1969, anderson_itoh1975}.
Due to an unknown trigger, the two components can momentarily recouple (probably due to a mechanism known as vortex-mediated mutual friction \citep{andersson+2006,antonelli_haskell2020}): the transfer of angular momentum from the neutron superfluid to the normal component results in a transient spin-up of the observable component, giving rise to a glitch.

Glitching behaviour can be very different from pulsar to pulsar \cite{mckenna_lyne1990,lyne+2000,espinoza+2011}, and its information can be encoded with a study of the glitch size and waiting time distributions (see, e.g., \cite{melatos+2008, howitt+2018, fuentes+2019}); however, the identification of precise trends is difficult due to the scarcity of data for some objects: because of a combination of intrinsic physical properties and the different time spans of observations, some objects have presented only one glitch, while some others have displayed a statistically more relevant number of events (up to 45 in PSR J0537-6910 \cite{antonopoulou+2018} and 23 in Vela \cite{espinoza2020arXiv}).

Some information about the structure of a glitching pulsar and the processes regulating this phenomenon can be obtained from its glitching behaviour, like the largest displayed glitch \cite{antonelli+2018} or the short-time angular velocity evolution after a glitch \cite{ashton+2019, pizzochero+2020, sourie_chamel2020, montoli_magistrelli+2020}.
Furthermore, it is possible also to obtain information about the neutron star structure (in particular, on the ratio between the moments of inertia of the normal and superfluid components \cite{datta_alpar1993,link+1999,andersson+2012,chamel2013,delsate+2016}) from the observed activity parameter (i.e., the average spin-up due to glitches; \mbox{see, e.g., \cite{mckenna_lyne1990})} of a pulsar.
This parameter is particularly interesting as pulsar activity observations, together with theoretical modelling of the thermo-rotational evolution of a pulsar, have also been used to provide indirect mass estimates of isolated neutron stars \cite{ho+2015,pizzochero+2017,montoli+2020}.

Despite early models considered the superfluid neutrons in the core (see, e.g., \citep{baym+1969,packard71}), after the seminal work of \citet{anderson_itoh1975}, the superfluid participating in the glitch has been generally thought to be limited in the crust of the star, where vortex pinning is possible \cite{alpar1977ApJ,donati2004NuPhA,seveso+2016}.
In fact, pulsar activity measured in the Vela pulsar seemed at first to be compatible with this idea of a crust-limited superfluid reservoir \cite{datta_alpar1993,link+1999}.
However, the introduction in the model of entrainment---a non-dissipative interaction that couples the two components \cite{andreev_bashkin1976}---and the calculation of this parameter in the neutron star crust \cite{chamel2012} have posed serious issues to the modelling: entrainment coupling in the crust diminishes the effective angular momentum reservoir, making it difficult for a stellar model with a crust-limited reservoir to display a Vela-like activity.

Currently, to justify the observed activity of Vela, the neutron star crust must be sufficiently thick to store a significant amount of angular momentum, corresponding
to a fraction of crust moment of inertia in a range going from 1.6 \% up to $\sim$10\%, depending on the importance of the effect of crustal entrainment, which is
currently under \mbox{debate \citep{sauls+2020}.}
Alternatively, some models \cite{guerci_alpar_2014,ho+2015,montoli+2020,sourie_chamel2020} also consider the possibility of a superfluid angular momentum that extends in the outer core: quantised vortex lines in the core superfluid could pin against the quantised flux lines of the proton superconductor (\cite{Muslimov1985,Srinivasan1990,drummond2017MNRAS}; see \mbox{also \citep{Alpar_pinningCore_2017}} for a review), so that it would be possible to store angular momentum in a region that is not just confined within the inner crust. However, there are some uncertainties about the nature of the proton superconductor \cite{Wood2020arXiv}, as well as the nature of neutron vortices in the outer core \cite{LeinsonMNRAS2020}, so that pinning with flux tubes is quite uncertain to date.

Therefore, a reliable estimation of the glitch activity and its associated uncertainty is crucial to validate the crustal origin of pulsar glitches. The problem is particularly interesting for those pulsars that do not show a clear linear relation between the cumulative glitch size and the observational time. 
In this paper, we will deal with this problem, trying to find new ways to calculate glitch activity and in particular its uncertainty, stressing some subtleties regarding the latter value.
Finally, we will employ the calculated activity parameter in a revised version of the original argument for the moment of inertia constraint found in \cite{link+1999, andersson+2012, chamel2013}.

\section{Extracting the Activity Parameter from Observations}
\label{sec:activity_calculation}

We consider a certain pulsar that has undergone $N_\textrm{gl}$ glitches with size $\Delta \Omega_i$ (with $i = 0, \ldots, N_\textrm{gl}-1$) in a long observational time interval $T_\textrm{obs}$. The absolute activity of a pulsar can be defined as
\begin{equation}
	\mathcal{A}_a = \frac{1}{T_\textrm{obs}} \sum_{i=0}^{N_\textrm{gl}-1} \Delta \Omega_i \, . 
\label{eq:aa}
\end{equation}
Strictly speaking, this definition of $	\mathcal{A}_a$ refers to a particular time window $T_\textrm{obs}$. However, if the rate in \eqref{eq:aa} is almost constant when restricted to shorter time windows within $T_\textrm{obs}$ (stationarity hypothesis), then a unique activity $\mathcal{A}_a$ can be defined for a long period, and \eqref{eq:aa} provides a reliable estimator for it (see, e.g., \cite{corral2006} for an analogous discussion on the mean seismic rate of earthquakes in a given time window). 
It is useful to introduce also the dimensionless activity parameter $\mathcal{G}$, defined as
(see, e.g.,~\cite{chamel2013})
\begin{equation}
	\mathcal{G} = {|\dot \Omega_\infty|}^{-1} {\mathcal{A}_a} \, ,
	\label{eq:G}
\end{equation}
where $|\dot{\Omega}_\infty|$ is the absolute value of the secular spin-down rate of the pulsar, namely the average angular velocity derivative in the period $T_\textrm{obs}$ containing several glitches.
The variable $\mathcal{G}$ gives us an idea of the amount of spin-down reversed by glitches, and it allows for a better comparison of pulsars with different spin-down rates~\cite{montoli+2020}.

From the practical point of view, to use \eqref{eq:aa} with real data, the basic requirement is that the pulsar should regularly be monitored during the interval $T_\textrm{obs}$, without missing any glitch. This is in general not the case, but while large glitches can be easily detected, very small glitches, easier to miss, should not contribute to the activity in a significant way (unless they are extremely frequent). 
Furthermore, it is not always clear whether the duration $T_\textrm{obs}$ of the observational campaign has been long enough, so that $\mathcal{A}_a $ calculated via \eqref{eq:aa} really reflects the true activity of the pulsar under study. For this reason, we consider \eqref{eq:aa} as a theoretical definition of the true activity of a pulsar, in the limit of very long $T_\textrm{obs}$. In the following, we will explore how to extract estimates of $\mathcal{A}_a$ from real data.

\subsection{Ordinary Linear Regression on the Cumulative Glitch History}

\begin{figure} 
%\widefigure
\centering
	\includegraphics[width = 0.8\textwidth]{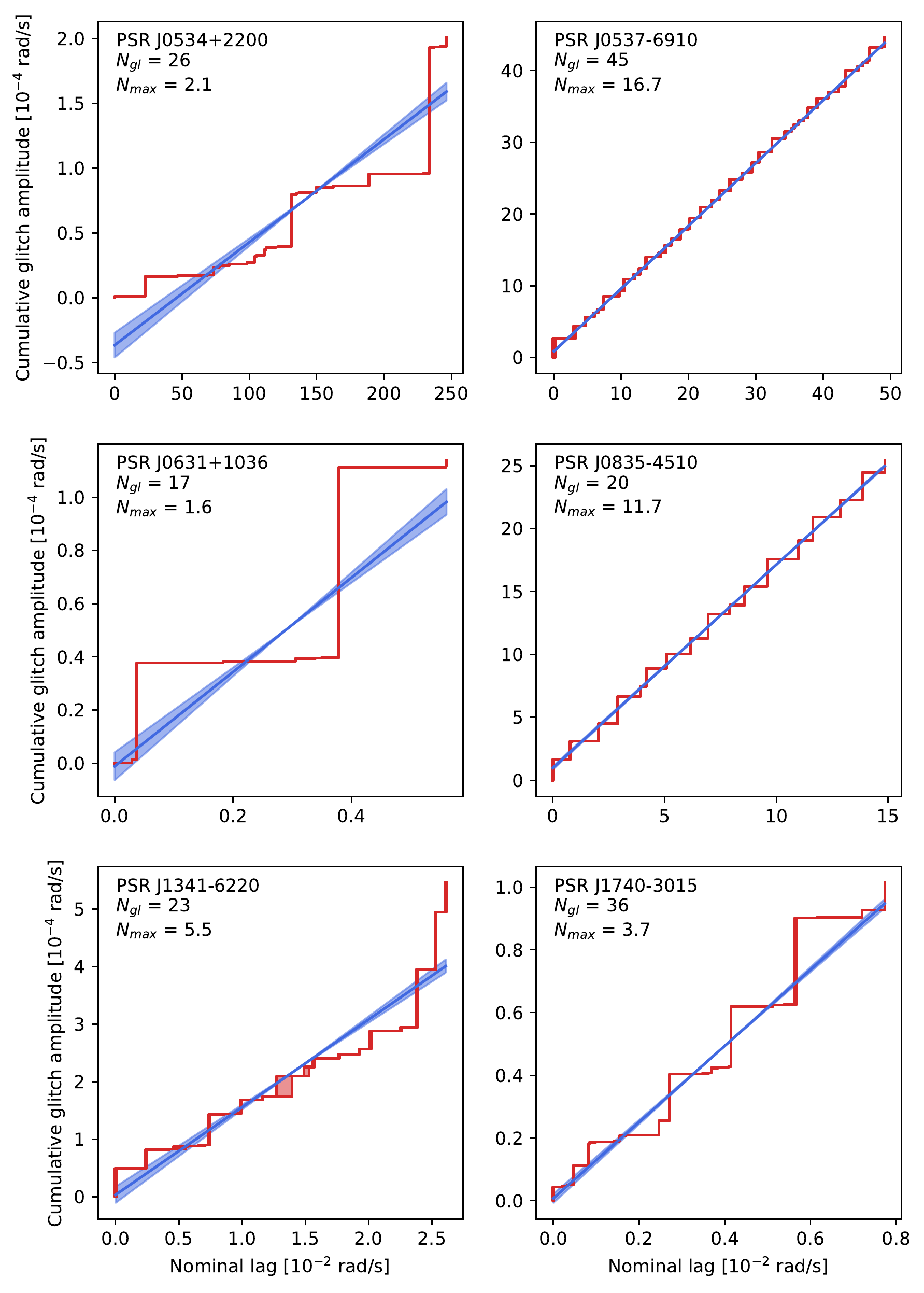}
	\caption{
	The glitch steps drawn by the cumulative points $(|\dot{\Omega}_\infty| t_i,\Omega_i)$, defined in \eqref{eq:cumul}, for the six pulsars with the largest number of glitches $N_{gl}$, as reported by the Jodrell Bank Glitch Catalogue \citep{espinoza+2011}.
	The activity is calculated with a least-squares linear fit on the midpoints of the cumulative glitch sequence (see \citep{wong+2001}), giving the blue curve with the associated uncertainty on the slope.
	Following Montoli et al. \citep{montoli+2020}, the plots are made by using the ``nominal lag'' $t |\dot{\Omega}_\infty|$ on the horizontal axis (a~rescaling of time $t$, which gives a rough estimate of the typical angular velocity lag accumulated between the spinning down normal component and a pinned superfluid component). In this way, the slope of the blue curves is the dimensionless activity $\mathcal{G}$.
	}
	\label{fig:activity_fit}%in the figures throughout, use () not [] around units
\end{figure}

Let us assume that the information at our disposal consists only of a list of glitch dates $t_i$ and amplitudes $\Delta\Omega_i$. For simplicity, let us neglect the extra information that is possibly contained in the value of $T_\textrm{obs}$, but note that $T_\textrm{obs}>t_{N_\textrm{gl}-1}-t_0$. 
Under this assumption, the absolute activity is usually calculated by fitting the cumulative glitch amplitude (\mbox{see, e.g., \citep{link+1999,wong+2001}})%please check the layout for the citations
. In this case, the relationship between angular velocity and time is described by the equation
\begin{equation}
 \Omega_i = \mathcal{A}_a \, t_i + q + \varepsilon_i\, ,
 \label{eq:ols}
\end{equation}
where $q$ is the vertical intercept and $\varepsilon_i$ are independent random variables with zero expectation and the same variance (homoscedasticity).
In the above formula, $\Omega_i$ and $t_i$ represent the angular velocity acquired by the star due to glitches and the time passed since the \mbox{first glitch, }
\begin{equation}
 \Omega_i = \sum_{j=1}^i \Delta\Omega_j 
 \qquad \qquad 
 t_i = \sum_{j=1}^i \Delta t_j \, ,
 \label{eq:cumul}
\end{equation}
where $\Delta t_j$ is the waiting time preceding the $j$-th glitch. This procedure sacrifices the information relative to the first glitch amplitude, $\Delta\Omega_0$, as the slope of the points in \eqref{eq:cumul} does not change for vertical translations.
One possibility to partially solve this issue is that of fitting the midpoints of the glitch steps drawn by the cumulative points $(t_i \, , \, \Omega_i )$, instead of the points themselves \cite{wong+2001,montoli+2020}.
An example of the activity fit performed using this prescription is shown in Figure~\ref{fig:activity_fit} for the six pulsars that have displayed the largest number of glitches $N_{gl}$ at the time of writing: the fit seems to capture the average slope of the glitch series, at least for J0537-6910 and Vela.
All the glitch sizes and waiting times employed in this paper were retrieved from the Jodrell Bank Glitch Catalogue\footnote{{\url{http://www.jb.man.ac.uk/pulsar/glitches/gTable.html}}} \citep{espinoza+2011}, while the spin-down rates of the stars from the ATNF%define if appropriate
 Pulsar Catalogue\footnote{{\url{https://www.atnf.csiro.au/research/pulsar/psrcat/}}} \citep{manchester+2005}.

 The central assumption behind this standard linear regression procedure is that the statistical properties of the processes underlying the glitch behaviour should not change if the window of observation is translated in time (i.e., the glitch series observed in a pulsar may be modelled as the outcome of a stationary stochastic process in the long run \cite{fulgenzi+2017}). 
If this is the case, then the available data set should correspond to a stationary sequence of glitches where possible aftershocks and more quiet periods of activity are both present many times, intertwined in such a way to produce an overall stationary spin-up rate (eventually, also many very small and undetected glitches may be included, as their contribution to the cumulative glitch amplitude is negligible). In this way, the activity calculated on sub-intervals should fluctuate around the asymptotic value calculated by considering an interval $T_\textrm{obs}$ containing several glitches \cite{corral2006}.

%\newpage
%
%\end{paracol}
%\nointerlineskip 
%\begin{paracol}{2}
%\linenumbers

For some pulsars, however, the observational window is so limited and the number of glitches detected so small that it may be unsafe to conclude that the observed value of $\mathcal{A}_a$ corresponds to the value that would be extrapolated by looking at a longer sequence of glitches. 
Practically, to be able to perform a standard linear regression
on the cumulative data, we have to demand that the available data fulfil two practical requirements \cite{montoli+2020}.
Firstly, the number of glitches should be significant, say $N_\textrm{gl} > 3$.
Secondly, at least two glitches of size comparable to $\Delta \Omega_\textrm{max}$, the maximum glitch amplitude in the sequence, should be present: a linear regression would poorly fit the set of data if the largest glitch $\Omega_\textrm{max}$ is significantly larger than all the others.
This last property can be quantified by demanding that the parameter $N_\textrm{max}$, defined as \cite{pizzochero+2017,montoli+2020}:
\begin{equation}
	N_\textrm{max} \, =\, \Delta \Omega_\textrm{max}^{-1} \, \sum_{i = 0}^{N_\textrm{gl}-1} \Delta \Omega_i
	> 1\, \quad \text{where} \qquad
	\Delta \Omega_\textrm{max} = \max_{i =0, \ldots, N_\textrm{gl}-1} \, \Delta \Omega_i \, ,
	\label{eq:Nm}
\end{equation}
be larger than $\sim$2 \cite{montoli+2020}.
However, in view of the hypothesis of stationarity, even if the glitches respect the above conditions, the ordinary least-squares linear regression may not be a well justified method, at least from the theoretical point of view. 
In fact, the technical point of how to extract $\mathcal{A}_a$ and the associated uncertainty, especially in the case of small $N_\textrm{gl}$ or $N_\textrm{max}$, have not been discussed in detail yet.

\subsection{Linear Regression on Heteroscedastic Data}

Besides all the issues related to linear regression mentioned in the previous section, there is one more linked to the fact that the data points in \eqref{eq:cumul} are not independent, as they arise from a cumulative construction.
Hence, the resulting residuals (which have to be minimised in the standard regression procedure) may not be independent and identically distributed: 
the fit residuals will not have the same variance, so it is no longer possible to assume homoscedasticity, which is a basic assumption of ordinary linear regression.

For this reason, we present here an alternative way to calculate the activity, but this time relaxing the hypothesis of homoscedasticity, by following a procedure for fitting cumulative data discussed by \citet{mandel1957}.
Let us assume that waiting times and glitch sizes are related by
\begin{equation}
 \Delta \Omega_i = \mathcal{A}_a\, \Delta t_i + \varepsilon_i\, .
 \label{eq:lin_relation}
\end{equation}
Note that the above equation may not be true in general, as there seems not to be a correlation between glitch sizes and waiting times \cite{melatos+2018}.
Using Equation \eqref{eq:cumul}, the cumulative times and sizes follow a relation:
\begin{equation}
 \Omega_i = \mathcal{A}_a\, t_i + \sum_{j = 1}^i \varepsilon_j \, ,
\end{equation}
which justifies our assumption of heteroscedasticity, as the deviations $\sum_{j=1}^i \varepsilon_j$ have different variance for each $i$.
It can be shown that the best unbiased linear estimator is given by \cite{mandel1957}
\begin{equation}
 \mathcal{A}_a = 
 \frac{\sum_{i}{\Delta \Omega_i}}{\sum_{i}{\Delta t_i}} 
 = 
 \frac{\sum_{i}{\Delta \Omega_i}}{t_{N_\textrm{gl}-1}-t_0} 
 \qquad\quad i=1,...,N_\textrm{gl}-1
 \label{eq:act_def}
\end{equation}
which is similar to the definition in~\eqref{eq:aa}.
The variance of this value is given by \cite{mandel1957}
\begin{equation}
 \mathrm{Var}(\mathcal{A}_a) = \frac{1}{(N_\textrm{gl}-1) (t_{N_\textrm{gl}-1}-t_0) } 
 \, \sum_{i} \frac{(\Delta \Omega_i - \mathcal{A}_a \Delta t_i)^2}{\Delta t_i} \, 
 \qquad\quad i=1,...,N_\textrm{gl}-1
\end{equation}
We present in Table~\ref{tab:activities} the best estimator of the dimensionless activity $\mathcal{G}$, with its standard deviation obtained with this method.
This value was obtained by neglecting the first glitch size $\Delta\Omega_0$ in each pulsar sequence, in order to have the same number of glitch sizes and preceding waiting times.
The most interesting feature we notice from the results is that the standard deviation when we assume homoscedasticity is almost one order of magnitude smaller than that with heteroscedasticity.

\begin{table} 
\centering
\small
	\caption{Dimensionless activities and their standard deviations, calculated for the six pulsars with the largest number of glitches, with a least-squares linear fit on the cumulative midpoints assuming homoscedasticity ($\mathcal{G}_\textrm{hom}$), with a least-squares linear fit assuming heteroscedasticity ($\mathcal{G}_\textrm{het}$), with a bootstrap on the size and waiting time samples separately ($\mathcal{G}_\textrm{rand}$), and on the pairs' size, preceding waiting time ($\mathcal{G}_\textrm{pre}$), and size, following waiting time ($\mathcal{G}_\textrm{post}$).}
	\label{tab:activities}
\setlength{\tabcolsep}{2.4mm}{ \begin{tabular}{llllll}
 \toprule
 \textbf{Pulsar} & \boldmath{$\mathcal{G}_\textrm{\textbf{hom}} (\%)$} & \boldmath{$\mathcal{G}_\textrm{\textbf{het}} (\%)$} & \boldmath{$\mathcal{\textbf{G}}_\textrm{\textbf{rand}} (\%)$} & \boldmath{$\mathcal{\textbf{G}}_\textrm{\textbf{pre}} (\%)$} & \boldmath{$\mathcal{G}_\textrm{\textbf{post}} (\%)$} \\
 \midrule
 0534+2200 & 0.0079 $\pm$ 0.0007 & 0.008 $\pm$ 0.006 & 0.008 $\pm$ 0.005 & 0.008 $\pm$ 0.005 & 0.008 $\pm$ 0.005 \\ 
 0537-6910 & 0.874 $\pm$ 0.003 & 0.85 $\pm$ 0.15 & 0.89 $\pm$ 0.11 & 0.86 $\pm$ 0.11 & 0.88 $\pm$ 0.03 \\
 0631+1036 & 1.77 $\pm$ 0.18 & 2.03 $\pm$ 1.95 & 2.11 $\pm$ 1.67 & 2.29 $\pm$ 1.80 & 1.80 $\pm$ 0.85 \\
 0835-4510 & 1.62 $\pm$ 0.02 & 1.6 $\pm$ 0.2 & 1.65 $\pm$ 0.3 & 1.6 $\pm$ 0.2 & 1.6 $\pm$ 0.2\\
 1341-6220 & 1.52 $\pm$ 0.10 & 1.9 $\pm$ 0.6 & 2.0 $\pm$ 0.6 & 1.9 $\pm$ 0.6 & 1.9 $\pm$ 0.5\\
 1740-3015 & 1.22 $\pm$ 0.04 & 1.3 $\pm$ 0.7 & 1.3 $\pm$ 0.5 & 1.3 $\pm$ 0.5 & 1.2 $\pm$ 0.45\\
 \bottomrule
 \end{tabular}}
\end{table}

\subsection{Extracting the Activity from Glitch Size and Waiting Time Distributions}

One alternative way to solve the issue in the calculation of the activity with the linear fit requires employing the probability distributions for the waiting times and sizes of the glitches of a particular pulsar.
In this way, it is possible to relax the hypothesis of linear dependence between waiting times and glitch sizes.

Probability distributions of glitch sizes and waiting times have been obtained in several previous works \cite{melatos+2008, howitt+2018, fuentes+2019}, which show that---as a general trend---glitch sizes seem to be consistent with a power-law distribution, while the waiting times are consistent with an exponential distribution. Vela and PSR J0537-6910 are somehow exceptional, as they seem be well described by a normal distribution in both size and waiting time.

Starting from the probability distribution of the waiting times $\Delta t$ and sizes $\Delta \Omega$, it is possible to infer some information about the probability distribution $P_{\mathcal{A}_N}$ for the activity parameter $\mathcal{A}_N$ after $N$ glitches.
Note that $\mathcal{A}_N$ and $\mathcal{A}_M$, for $N\neq M$, are different random variables, distributed according to different laws, i.e., $P_{\mathcal{A}_N} \neq P_{\mathcal{A}_M}$. 

Given the definition of activity, its distribution can be obtained by considering the ratio of two random variables, the sum of sizes $\Delta \tilde{\Omega}_N$ and of waiting times $\Delta \tilde{t}_N$. 
The latter, in turn, are the sum of random variables themselves, i.e., the single glitch size $\Delta \Omega_i$ and the single waiting time $\Delta t_i$, so that their densities $P_{\Delta \tilde{\Omega}_N}$ and $P_{\Delta \tilde{t}_{N}}$ can be obtained by means of repeated convolutions,
\begin{align}
	\Delta \tilde{\Omega}_N = \sum_{i=1}^N \Delta \Omega_i \qquad \Rightarrow& \qquad \Delta \tilde{\Omega}_N \sim P_{\Delta \tilde{\Omega}_N} = \underbrace{P_{\Delta \Omega} * \ldots * P_{\Delta \Omega}}_{N \text{ times}}
	\label{eq:rV_s}\\
	\Delta \tilde{t}_N = \sum_{i=1}^N \Delta t_i \qquad\Rightarrow& \qquad \Delta \tilde{t}_{N} \sim P_{\Delta \tilde{t}_{N}} = \underbrace{P_{\Delta t} * \ldots * P_{\Delta t}}_{N \text{ times}}
	\label{eq:rV_t}
\end{align}
Since $\mathcal{A}_N=\Delta \tilde{\Omega}_N/\Delta \tilde{t}_{N}$, its distribution can be obtained through
\begin{equation}
	P_{\mathcal{A}_N}(a) = 
	\int_{-\infty}^{\infty} \mathrm{d}x\, |x|\, P_{\Delta \tilde{t}_{N}}(x)\, P_{\Delta \tilde{\Omega}_{N}}(x a)\, .
\end{equation}
The main advantage of this method is that, with only the assumption of independence between sizes and waiting times in the stationary regime, it is possible to obtain the whole probability distribution of~$\mathcal{A}_N$.

Although this is a possible method to extract $\mathcal{A}_N$, it is troublesome to numerically obtain this distribution, as a convolution of $N$ probability distributions, albeit identical, starts to be infeasible when $N$ becomes large.
The calculation can be simplified by obtaining the first two moments of $P_{\mathcal{A}_N}$, the mean and the variance, by employing a generalised version of the central limit theorem, the delta method (see Appendix \ref{app:delta}).
This method allows us to calculate the mean and the variance of $P_{\mathcal{A}_N}$ starting from the mean and variance of the two distributions $P_{\Delta \Omega}$ and $P_{\Delta t}$.
This procedure, however, is problematic if $P_{\Delta t}$ or $P_{\Delta \Omega}$ have a non-well-defined variance.

\subsection{Estimating the Uncertainty of Activity: The Bootstrap Method}

An attractive alternative to the methods described above, which does not build on the linear assumption in \eqref{eq:lin_relation}, is the so-called ``bootstrap method'' \cite{bootstrap}.
The idea is that of resampling with replacement the original data in order to calculate some statistics, as, e.g., the mean and standard deviation of the calculated activity.
In our case, the samples are two: the list of the waiting times $\Delta t_i$ (of length $N_\textrm{gl}-1$) and the list of sizes $\Delta\Omega_i$ (of length $N_\textrm{gl}$).
Of course, we have to draw the same number ($N_\textrm{gl}-1$) of waiting time-size couples in order to have a fair estimation of the activity and its standard deviation.
To avoid the homoscedasticity problem, the pulsar activity is calculated by employing the definition in Equation~\eqref{eq:act_def} on each set of resampled data.
We can also take into account the possibility of a dependency between a glitch size and the preceding or the subsequent waiting time, so it is useful to also bootstrap on the other two samples: the sample made up by ordered pairs $\{ (\Delta \Omega_i, \Delta t^{\textrm{pre}}_{i}) \}_{i = 1, \ldots, N_\textrm{gl}-1}$, where $\Delta t^{\textrm{pre}}_{i}$ is the waiting time preceding the glitch of size $\Delta \Omega_i$, and the sample comprised by ordered pairs $\{(\Delta \Omega_i, \Delta t^{\textrm{post}}_{i})\}_{i = 0, \ldots, N_\textrm{gl}-2}$, where $\Delta t^{\textrm{post}}_{i}$ is the waiting time following a glitch of size $\Delta \Omega_i$.

\begin{figure} 
%\widefigure
\centering
	\includegraphics[width= .7 \textwidth]{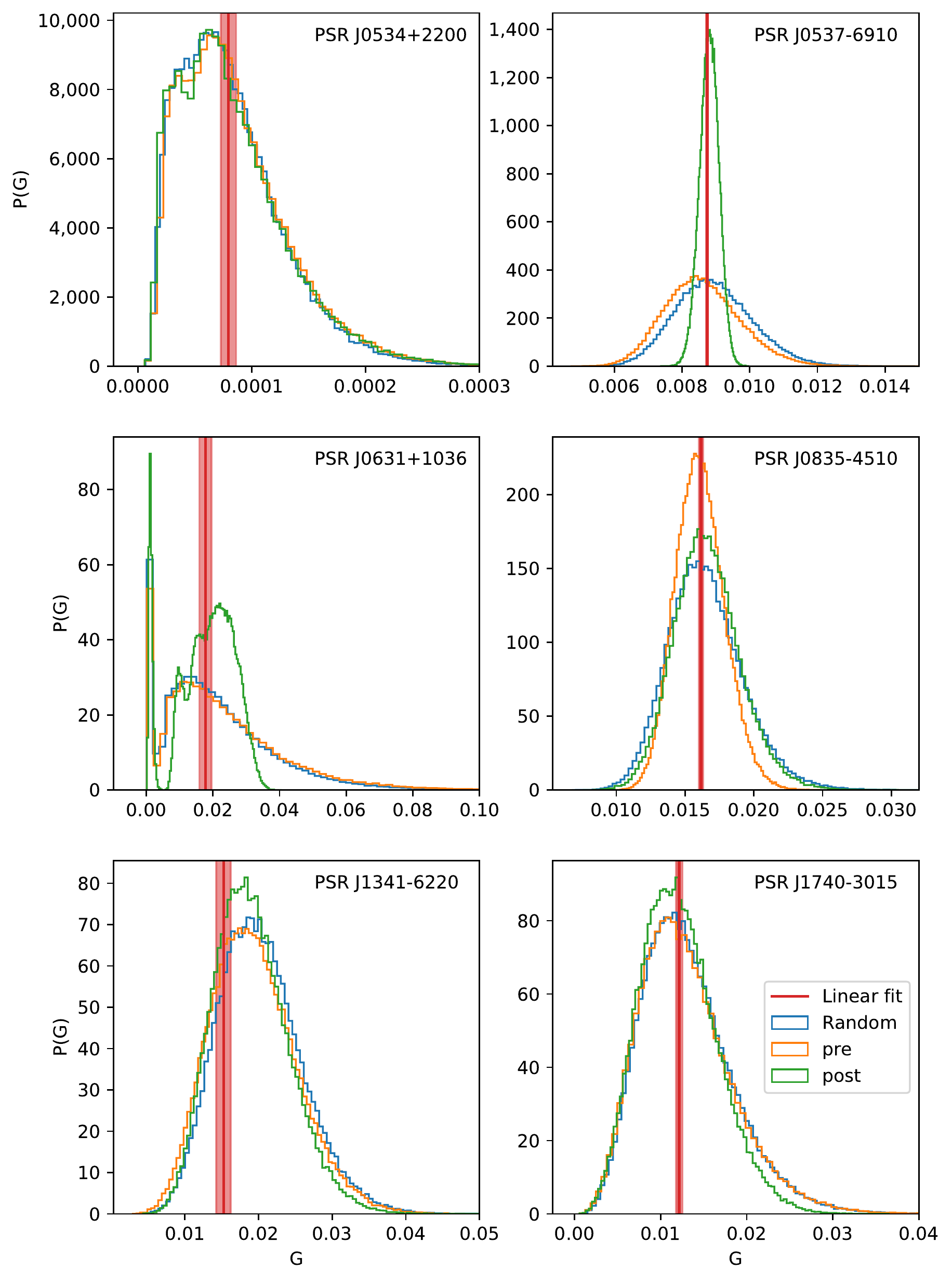}
	\caption{Dimensionless pulsar activity $\mathcal{G}$, calculated by sampling both the size and waiting time samples randomly (in blue), by sampling the pair $(\Delta \Omega,\Delta t_\textrm{pre})$ (in orange) and the pair $(\Delta \Omega,\Delta t_\textrm{post})$ (in green).
	The results of the linear fit of the cumulative glitch data are also plotted (in red; the shaded area is the $1\sigma$ region).}
	\label{fig:bootstrap}
\end{figure}

Figure~\ref{fig:bootstrap} shows the histograms obtained by resampling the data $10^4$ times in all three cases described above.
In the same plot, also the dimensionless activities obtained as a result of the ordinary linear regression on the cumulative glitch data are displayed.
We can see that the activity calculated by means of bootstrapping is compatible with the results obtained from an ordinary linear regression, but it generally has larger standard deviations (see also Table~\ref{tab:activities}).
It is interesting to notice the case of PSR J0537-6910, one of the few stars that presents a significant correlation between the glitch size and the following waiting time \cite{middleditch+2006}.
This correlation shows its effects also in Figure~\ref{fig:bootstrap}: the histogram for J0537-6910, in the particular case of the sample of size-following waiting time pairs, is much more peaked than the other two cases.
At lower confidence, also PSR J0631+1036 shows a correlation between size and the following waiting time and Vela a correlation between size and the preceding waiting time \citep{melatos+2018}. These correlations show their effect in the histograms as well.

It is also interesting to notice the peculiar form of the PSR J0631+1036 activity distribution: it shows two clear peaks, one on very small values and one around $\mathcal{G} \approx 0.02$.
This is probably because of the particular glitch sequence of this star (see Figure~\ref{fig:activity_fit}): it displays two very large glitches and many others with sizes several orders of magnitude smaller.
Thus, it is likely that the peak on smaller values was generated by sampling the small glitches only, while the peak on larger values occurs when one or both large glitches were sampled.
Moreover, as a consequence of the particular glitch sequence for this star, the value of $N_\textrm{max}$ for this star is smaller than the ones of the other stars in the sample: this value may increase in the future, by observing large glitches, giving us a more reliable estimate of the \mbox{activity \cite{montoli+2020}.}

\begin{figure} 
%\widefigure
\centering
	\includegraphics[width = 0.87 \textwidth]{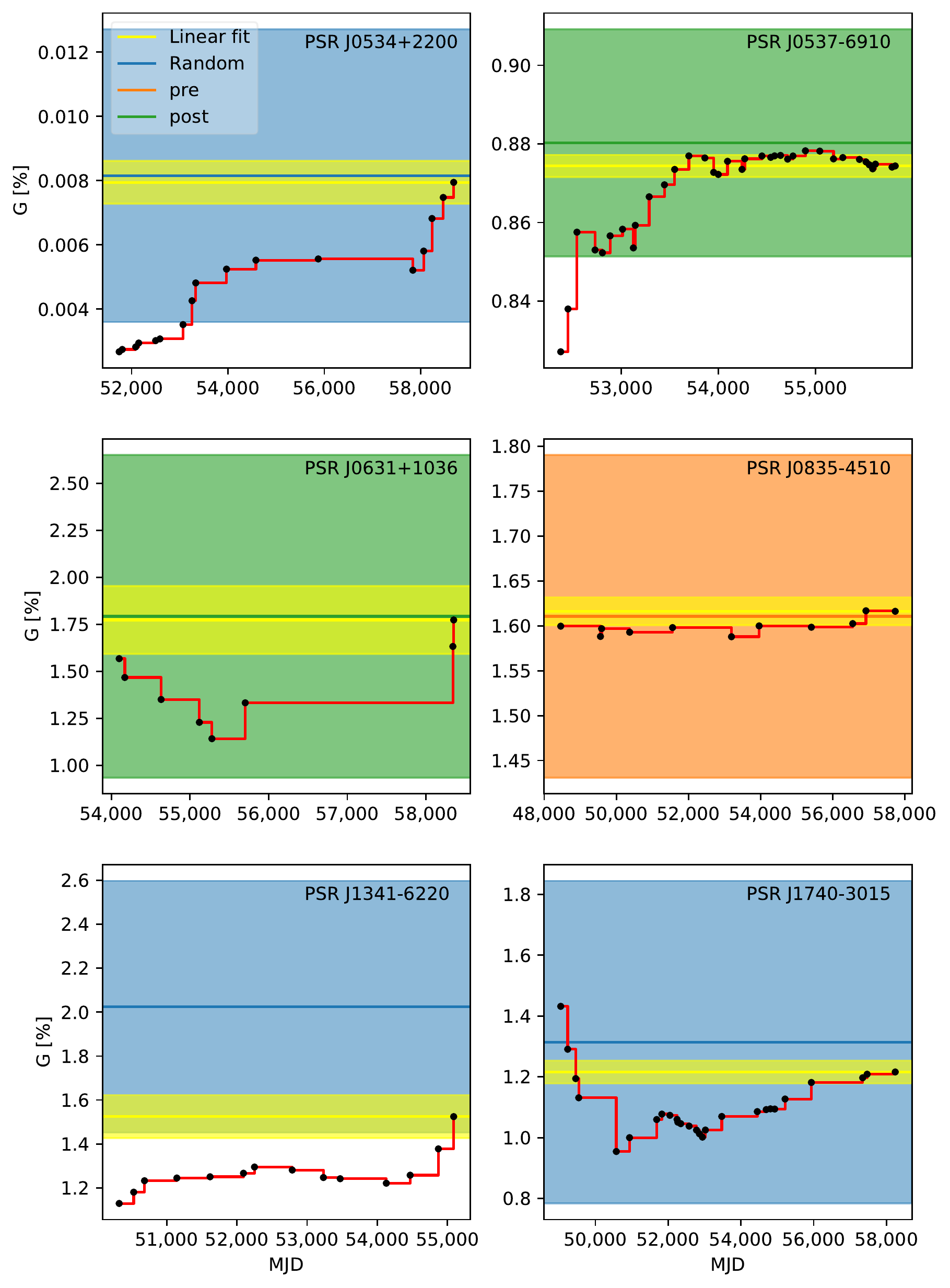}
	\caption{Evolution of the glitch activity over time.
	The first point of each curve is the activity calculated employing the linear fit on the cumulative midpoints of the first ten glitches, assuming homoscedasticity.
	The subsequent points are calculated by gradually adding all the glitches that pulsar displayed.
	For each pulsar, the estimate of the activity via ordinary linear regression with all glitches (in yellow) is shown, along with its uncertainty (shaded).
	We also present the bootstrap estimates and their uncertainties in the $\mathcal{G}_\textrm{post}$ case for PSR J0537-6910 and PSR J0631+1036 (green), $\mathcal{G}_\textrm{pre}$ for PSR J0835-4510 (Vela, orange) and $\mathcal{G}_\textrm{rand}$ for all other stars (blue).}
	\label{fig:gradual_activity}
\end{figure}

In Figure~\ref{fig:gradual_activity}, we try to give an idea of how much the activity changes when a new glitch occurs.
The first point of each curve is the activity calculated with the linear fit on the cumulative midpoints using the first ten glitches, assuming homoscedastic data. 
Then, we update the activity value with the same method whenever a new glitch is displayed.
We also plot the activity parameter calculated with the ordinary linear regression and all the glitches, along with its uncertainty.
For PSR J0537-6910 and PSR J0631+1036, we present the bootstrap estimate $\mathcal{G}_\textrm{post}$ and its standard deviation; for Vela, we show the same with $\mathcal{G}_\textrm{pre}$; while for all the other pulsars, we present the case with uncorrelated glitch sizes and waiting times ($\mathcal{G}_\textrm{rand}$).
The idea is that of taking into account the size-waiting time correlations, where present.
We note that, except the very particular cases of Vela and PRS J0537-6910, the activity evolution of each star generally lies outside the error region for the linear fit with homoscedastic data for all the glitch history, except---of course---the latest glitch.
The bootstrap uncertainty better describes the variance of the glitch history.
A notable exception is that of PSR J1341-6220, which is well below the error bar for both the activity calculations, except for the last three glitches.
This is because these glitches are three of the largest ones displayed by this pulsar (see also Figure~\ref{fig:activity_fit}).
In general, however, it is interesting to notice how variable the activity parameter is.
A single large glitch can change its value (see, e.g., the Crab pulsar after its November 2017 glitch \cite{shaw+2018}).
This fact stresses the importance of having a much larger uncertainty on the activity, which is the result of not assuming homoscedasticity in the linear fit or linear dependence between glitch sizes and waiting times.

%\newpage
%\end{paracol}
%\nointerlineskip 

%\begin{paracol}{2}
%\linenumbers
%\switchcolumn
%

\section{Moment of Inertia Constraint}
\label{sec:moi_constraint}

\textls[-15]{The activity parameter allows extracting information on the moment of inertia fraction of the superfluid reservoir in a glitching pulsar \cite{datta_alpar1993,link+1999}.
In this section, we present a revised version of the constraint on the moment of inertia relative to the superfluid component derived by Link et al. \citep{link+1999}; see also Andersson et al. \citep{andersson+2012} and Chamel \citep{chamel2013} for the inclusion of entrainment coupling between the normal and superfluid components. Our derivation is presented in Appendix~\ref{app:moi_constraint} and takes into account the non-rigid rotation of the superfluid component, the stellar stratification and the non-uniform entrainment coupling between the components. The constraint is given by Equation \eqref{cristobaldo}, which can be written in terms of the total moment of inertia $I$ and of the moment of inertia of the superfluid component $I_v$ as  }
\begin{equation}
	\frac{I_v}{I - I_v} > \mathcal{G} \, .
	\label{eq:act_constraint}
\end{equation}
Assuming that the superfluid reservoir extends in the layers between $R_c$ (the crust-core transition radius) and $R_d$ (the neutron drip radius), the value of $I_v$ in the slow-rotation approximation is  
\begin{equation}
	I_v = \frac{8\pi}{3 } \int_{R_c}^{R_d} dr \ r^4 e^{\Lambda(r)-\Phi(r)} \left[ \rho(r) + P(r) \right] \frac{y_n(r)}{1 - \varepsilon_n(r)} \frac{\Omega_p - \omega(r)}{\Omega_p}\, ,
	\label{eq:Iv}
\end{equation}
while the total moment of inertia $I$ is the usual one in the slow rotation framework \cite{hartle1967,andersson_comer2001,antonelli+2018}, given in Equation~\eqref{eq:I_rot}. 
In the above equation, $y_n$ is the superfluid neutron baryon density (limited to the region where pinning is possible) divided by the total baryon density, i.e., $y_n(r)$ is different from zero only where the superfluid can pin to inhomogeneities and maintain its state of motion while the normal component spins down: in this case, it is limited to the crust. 
The other quantities appearing in~\eqref{eq:Iv} are introduced in Appendix \ref{app:moi_constraint}, but it is important to remark here that the frame drag $\omega$ should contain a dependence on the angular velocities of both components \cite{andersson_comer2001,sourie+2017,antonelli+2018}, which was neglected in the derivation of~\eqref{eq:Iv}.

The first measurements of the activity parameter of the Vela pulsar and the moment of inertia fraction estimates for different equations of state (EoSs) seemed to be in accordance with the constraint~\eqref{eq:act_constraint} with~$\varepsilon_n=0$~\cite{link+1999}. 
Only later, the entrainment parameter~$\varepsilon_n$ in the crust of a neutron star was calculated in~\cite{chamel2012}, by estimating the effects of Bragg scattering on the superflow due to the presence of the crustal lattice.
These calculations yield a negative entrainment parameter $\varepsilon_n \sim -10 $ in a substantial portion of the inner crust, which implies a severely hindered motion of the superfluid component. 
This would reduce the amount of extra angular momentum stored in the crust between two glitches---and thus of $I_v$---making the requirement in \eqref{eq:act_constraint} more difficult to be met.
As a result of that, the only way for the star to acquire enough angular momentum between glitches to explain the observed activity is to have a large region inside it to store angular momentum (larger than the crust of the star) or to have an unreasonably small mass, around $\sim 0.5 \, M_\odot$.
This problem has been highlighted in several papers~\cite{andersson+2012, chamel2013, ho+2015, delsate+2016, carreau+2019}.

If we assume the superfluid region to be limited in the crust of the star and we fix the microphysical parameters, namely the EoS and the entrainment parameter, then the moment of inertia fraction in \eqref{eq:act_constraint} is a function of the mass of the star only.

\begin{figure} 
 \centering
 \includegraphics[width = 0.7\textwidth]{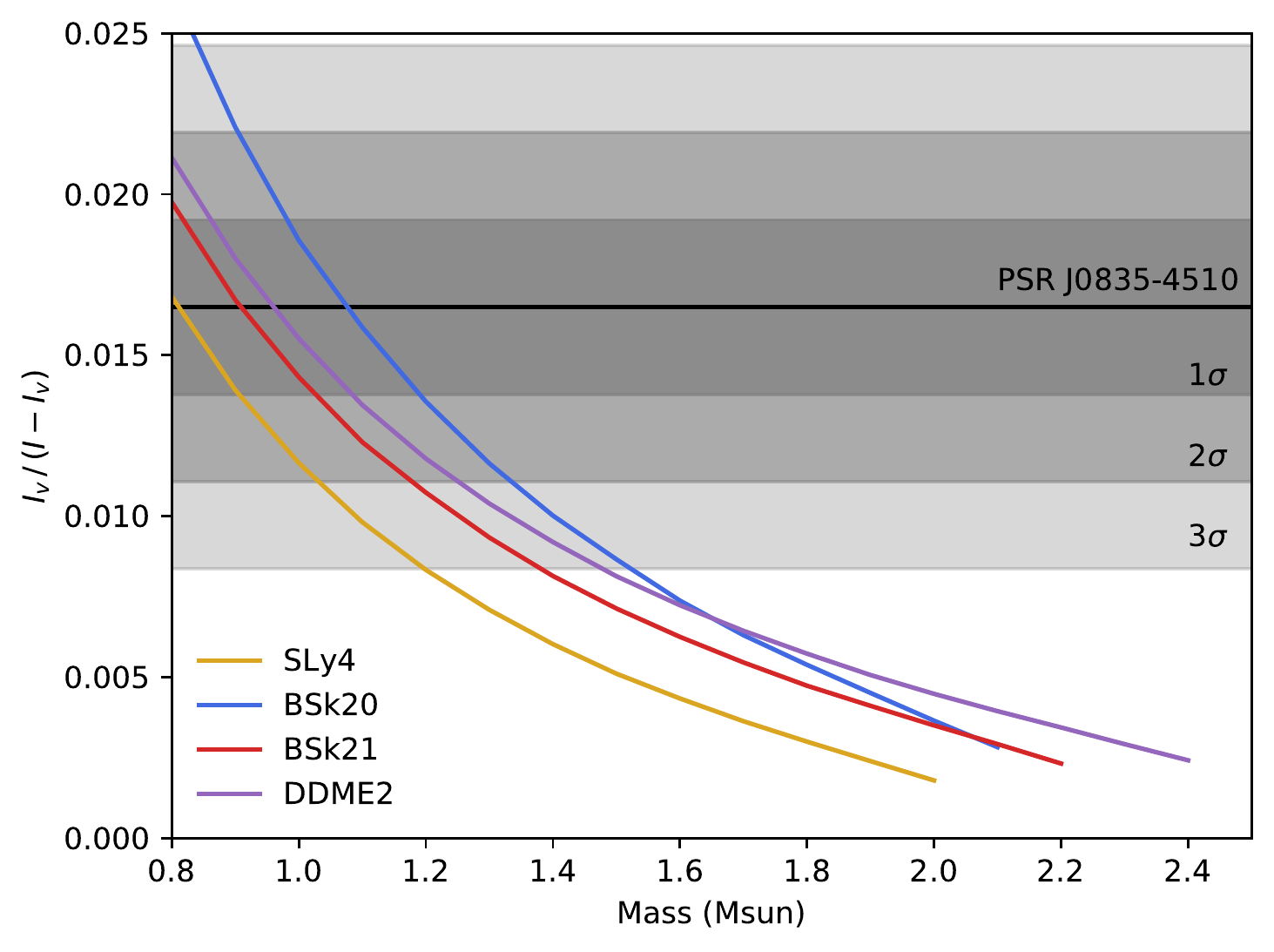}
 \caption{Activity constraint on the superfluid component moment of inertia plotted for some EoSs: SLy4 %define if appropriate
 \citep{douchin_haensel2001}, BSk20 %define if appropriate
 and BSk21 \citep{goriely+2010}, and the DDME2 %define if appropriate
 EoS \citep{lalazissis+2005}, glued with a SLy4 crust \citep[following the method described in][]{fortin+2016}. 
 The entrainment parameter is that calculated in \cite{chamel2012}, and the superfluid reservoir is limited in the crust of the star.
 The dimensionless activity parameter $\mathcal{G}$---calculated with the bootstrap method described in Section~\ref{sec:activity_calculation} (the case with random glitch sizes and waiting times)---is also plotted for the Vela pulsar, along with the $1\sigma$, $2\sigma$ and $3\sigma$ uncertainties.}
 \label{fig:thecrust}
\end{figure}

In Figure~\ref{fig:thecrust}, we plot the quantity $I_v/(I - I_v)$ appearing in \eqref{eq:act_constraint} as a function of the stellar mass for some EoSs and for the entrainment parameter calculated by \citet{chamel2012}, assuming a superfluid reservoir limited to the crust.
As we can see, the constraint \eqref{eq:act_constraint} imposes that the high $\mathcal{G}$ value of the Vela pulsar (PSR J0835-4510) can be explained only if the mass of Vela is small, ranging from $\approx 1.1 M_\odot$ for the BSk20%define if appropriate
 EoS to $\approx 0.8 M_\odot$ for the SLy4%define if appropriate
 EoS.
Let us however consider also the $1\sigma$ uncertainty region, calculated with the bootstrap method described in Section~\ref{sec:activity_calculation}, with random waiting times and glitch sizes. In this case for the BSk20 EoS, the Vela pulsar has an upper limit on the mass of about~$1.2 M_\odot$.
Note that this value is slightly above the minimum mass of a neutron star estimated from the calculations of core-collapse supernovae (i.e.,~$1.17\, M_\odot$; see \cite{suwa+2018}) and the smallest mass measured in a neutron star ($1.174 \pm 0.004\, M_\odot$, measured in PSR J0453+1559, \cite{martinez+2015}).
If we consider the $3\sigma$ uncertainty range, then we obtain a limit of $\approx 1.5 M_\odot$ for the same EoS.
It is thus clear that a careful estimation of the activity parameter and the associated error is crucial if one is to set strict quantitative constraints on the moment of inertia involved in glitches and constrain the location of the superfluid~reservoir.

A more careful estimate of the glitch activity parameter may thus play an important role in resolving the tension between strong entrainment and models with a crust-limited reservoir.
Several other effects have, of course, been suggested and are likely to play a role in this problem, including a maximally stiff EoS \citep{piekarewicz+2014}, a Bayesian analysis of the EoS uncertainty \citep{carreau+2019} or an extension of the region where the neutron superfluid participates in the glitch beyond the crust-core transition, based on the assumption that only the superfluid in the ${}^1S_0$ state participates in the glitch phenomenon and on an analysis of the temperature of the star \citep{ho+2015}.
On the other hand, also different calculations of the entrainment parameter have been proposed
(e.g., \citep{martin_urban2016,watanabe_pethick2017, sauls+2020}), which yield milder entrainment effects in the crust. 
However, even if the high activity of Vela may be described in terms of a purely crustal reservoir by assuming a weaker entrainment, an analysis of the 2016 Vela glitch points to the need to invoke the neutron superfluid also in the core of the star \citep{graber+2018,HaskellKhomenko18,pizzochero+2020}, independently of the presence of entrainment in the model \cite{montoli_magistrelli+2020}.

%%%%%%%%%%%%%%%%%%%%%%%%%%%%%%%%%%%%%%%%%%

\section{Conclusions}

In this paper, we discussed different ways of calculating the activity parameter in glitching pulsars.
The most commonly used one is that of performing an ordinary linear regression on the cumulative glitch data, which is justified if one considers the activity parameter as an intrinsic characteristic of a glitching pulsar, and thus inferable from a limited observation of its glitching behaviour (which, in general, may not be stationary).
While this last statement might be true, it is also true that a strong autocorrelation or correlation between glitch sizes and waiting times has very rarely been observed \cite{melatos+2018, carlin_melatos2019_autocorrelations}. 
Moreover, fitting the cumulative data may also affect the uncertainty calculated from a ordinary linear regression, as the hypothesis of the homoscedasticity of the data cannot be satisfied.
Therefore, the main consequence of including these assumptions in the fitting procedure (namely, the linear dependence of glitch sizes and waiting times and the homoscedasticity of the cumulative data) is an underestimation of the activity uncertainty. 
 This is an important point, as in some pulsars, an additional single glitch (or a sequence of a few glitches) can significantly affect its observed activity, especially for those objects with low $N_\textrm{max}$ \cite{montoli+2020}, like the Crab pulsar or PSR J0631+1036 (see Figure \ref{fig:gradual_activity}).

A first alternative to calculate the activity is the linear regression developed by\linebreak \citet{mandel1957}, where the hypothesis of the homoscedasticity of the data was relaxed: this methodology is justified by the fact that cumulative data are not homoscedastic, since the variance of the data should increase as data are cumulated.
As a result, the uncertainty of the activity parameter increases by about one order of magnitude with respect to that calculated with an ordinary linear regression.

As a second step, we also relaxed the hypothesis of linear dependence between glitch sizes and waiting times, by employing their probability distribution estimates.~While employing the size and waiting time distributions is arguably the best way to include the information provided by observations of the pulsar in the activity parameter, it is also true that a study of the probability distribution of the activity parameter is computationally very challenging.
Furthermore, an approximate version of this estimate, obtained by employing a more general version of the central limit theorem, leads to difficulties, due to the fat-tailed probability distribution of the glitch sizes.

We thus employed an alternative way to calculate the activity and its uncertainty, relaxing the hypothesis of linear dependence, by bootstrapping on the sizes and waiting time data sets and calculating the activity by summing sizes and waiting times and calculating the ratio.
Much larger uncertainties are obtained, on the same order of magnitude of the linear fit on heteroscedastic data.
 A qualitative explanation of why a bootstrap estimate yields a larger uncertainty than the ordinary linear regression is the fact that the glitch activity is dominated by a few large events, especially for pulsars with low $N_\textrm{max}/N_\textrm{gl} \lesssim 0.1$, which is again the case for the Crab pulsar and J0631+1036; see Table \ref{tab:activities}.

We then used these results to study a revised version of the constraint on the moment of inertia of the superfluid component and on the mass of the star \cite{link+1999, andersson+2012, chamel2013}.
While the result obtained in this way is essentially the same for the mean value of the activity (i.e.,~the Vela activity does not allow a crust-limited reservoir and strong entrainment), the larger uncertainty allows for models in which the crust is enough, i.e., models in which the superfluid reservoir is located entirely in the crust for credible values of the star mass. For example, a model of a $1.2 M_\odot$ neutron star described by the BSk20 EoS is within the $1\sigma$ uncertainty region of the activity that we calculate.

%%%%%%%%%%%%%%%%%%%%%%%%%%%%%%%%%%%%%%%%%%
\vspace{6pt} 
%%%%%%%%%%%%%%%%%%%%%%%%%%%%%%%%%%%%%%%%%%

%\authorcontributions{ Conceptualization, methodology, formal analysis, draft preparation, A.M. and M.A.; review and editing, A.M., M.A., B.H. and P.P.; supervision, B.H. and P.P. All authors read and agreed to the published version of the manuscript.}

%\funding{M.A. and B.H. acknowledge support from the Polish National Science Centre Grants SONATA BIS 2015/18/E/ST9/00577 and OPUS 2019/33/B/ST9/00942.}

%Please add: ``This research received no external funding'' or ``This research was funded by NAME OF FUNDER grant number XXX.'' and and ``The APC was funded by XXX''. Check carefully that the details given are accurate and use the standard spelling of funding agency names at \url{https://search.crossref.org/funding}, any errors may affect your future funding.}

%\dataavailability{The data analysed in this paper can be found on the ATNF Pulsar Catalogue (\url{https://www.atnf.csiro.au/research/pulsar/psrcat/}, see \cite{manchester+2005}) and on the Jodrell Bank Glitch Catalogue (\url{http://www.jb.man.ac.uk/pulsar/glitches/gTable.html} see \cite{espinoza+2011}).}  %MDPI: In this section, please provide details regarding where data supporting reported results can be found, including links to publicly archived datasets analyzed or generated during the study. Please refer to suggested Data Availability Statements in section “MDPI Research Data Policies” at \href{https://www.mdpi.com/ethics}{https://www.mdpi.com/ethics}. You might choose to exclude this statement if the study did not report any data.

\acknowledgments{
Partial support comes from PHAROS, COST Action CA16214. M.A. and B.H. acknowledge support from the Polish National Science Centre Grants SONATA BIS 2015/18/E/ST9/00577 and OPUS 2019/33/B/ST9/00942. 
The data analysed in this paper can be found on the ATNF Pulsar Catalogue (\url{https://www.atnf.csiro.au/research/pulsar/psrcat/}, see \cite{manchester+2005}) and on the Jodrell Bank Glitch Catalogue (\url{http://www.jb.man.ac.uk/pulsar/glitches/gTable.html} see \cite{espinoza+2011}).
We thank Crist\'obal M. Espinoza and Andrew Melatos for providing interesting comments.
}

%\conflictsofinterest{The authors declare no conflict of interest.}

%%%%%%%%%%%%%%%%%%%%%%%%%%%%%%%%%%%%%%%%%%
%% optional
\appendixtitles{yes} % Leave argument "no" if all appendix headings stay EMPTY (then no dot is printed after "Appendix A"). If the appendix sections contain a heading then change the argument to "yes".
%\appendixstart
\appendix

\section{Activity Calculation with the Delta Method}
\label{app:delta}

We use the delta method (a generalisation of the central limit theorem) to extract the mean and standard deviation of the activity parameter $\mathcal{A}_N$ after $N$ glitches, given a probability distribution of the sizes $P_{\Delta \Omega}$ and waiting times $P_{\Delta t}$ of a pulsar. The method works under the simplifying assumption that the sizes and the waiting times are independent and identically distributed random variables. 
Let us recall the two random variables $\Delta \tilde \Omega_N$ and $\Delta\tilde t_N$ defined in Equations \eqref{eq:rV_s} and \eqref{eq:rV_t}.
 The random variable $\mathcal{A}_N$ is defined as 
 \begin{equation}
 	\mathcal{A}_N \, = \, \Delta \tilde \Omega_N / \Delta\tilde t_N \, .
 	\label{eq:rV_AN}
 \end{equation}
 Its expectation value is given by ($a \in \mathbb{R}^+$ denotes the value assumed by $\mathcal{A}_N$)  
 \begin{equation}
 	\mathrm{E}[\mathcal{A}_N] = \int a \, P_{\mathcal{A}_N} (a)\, \mathrm{d} a \, 
 	= \int \frac{\Delta \tilde \Omega_N}{\Delta \tilde t_N} 
 	\, \prod_{i=1}^{N} \, P_{\Delta t}(\Delta t_i) \, \mathrm{d} \Delta t_i 
 	\, \prod_{j=1}^{N} \, P_{\Delta \Omega}(\Delta \Omega_j) \, \mathrm{d} \Delta \Omega_j\, \, .
 \end{equation}
 The independence of the variables is loosely justified by observing the small correlation and autocorrelation in glitch sizes and waiting times \cite{melatos+2018, carlin_melatos2019_autocorrelations}, while being identically distributed is a working assumption.
 The above equation boils down to  
 \begin{equation}
 	\mathrm{E}[\mathcal{A}_N] = \mathrm{E}[\Delta \tilde \Omega_N]\, \mathrm{E} \left[\frac{1}{\Delta \tilde t_N} \right]\, .
 	\label{eq:mean_AN}
 \end{equation}
 To calculate the variance of $\mathcal A_N$, let us first calculate, for the sizes $\Delta \tilde \Omega_N$:
 \begin{align}
 	\mathrm{E}(\Delta\tilde \Omega_N^2) &= N \mathrm{E}[\Delta \Omega^2] + N(N-1) \mathrm{E}[\Delta \Omega]^2\\
 	\mathrm{E}(\Delta\tilde \Omega_N)^2 &= N^2 \mathrm{E}[\Delta \Omega]^2\\
 	\mathrm{Var} [\Delta\tilde \Omega_N] &= N( \mathrm{E}[\Delta \Omega^2] - \mathrm{E}[\Delta \Omega]^2)\, = \, N \, \mathrm{Var} [\Delta \Omega ]\, .
 \end{align}
Now, using the above results, a simple direct calculation gives 
 \begin{equation}
 	\mathrm{Var} [\mathcal{A}_N] 
 	= N \, \mathrm{Var}[\Delta \Omega] \, \mathrm{E}\left[\frac{1}{\Delta\tilde t_N^2}\right] 
 	+ 
 	N^2 \, \mathrm{E}[\Delta \Omega]^2 \, \mathrm{Var}\left[\frac{1}{\Delta\tilde t_N} \right] 
 	\, .
 	\label{eq:var_AN}
 \end{equation}
Let us now assume that $P_{\Delta \tilde t_N}$ converges to a normal distribution with mean $N\theta$ and variance $N\sigma^2$ as $N\rightarrow \infty$.
In this case, we can employ the delta method, which tells us that any function $g$ of the random variable $\Delta \tilde t_N$ will be distributed normally with mean $g(N\theta)$ with variance $N[g'(N\theta) \, \sigma]^2$.
Note that this assumption is not true if glitch sizes are described by a power law distribution with non-defined variance (but should hold for those pulsars like PSR J0537-6910).
Given the definition of $\Delta\tilde{t}_N$, we have that $\theta = \mathbb{E}[\Delta t]$ and $\sigma^2 = \mathrm{Var}[\Delta t]$, and by considering the cases $g(\Delta \tilde t_N) = (\Delta \tilde t_N)^{-1}$ and $g(\Delta \tilde t_N) = (\Delta \tilde t_N)^{-2}$, we obtain 
 \begin{equation}
 	\mathrm{E}\left[\frac{1}{\Delta \tilde t_N} \right] 
 	\approx \frac{1}{ N \, \mathrm{E}[\Delta t]} \quad\qquad
 	\mathrm{E}\left[\frac{1}{\Delta \tilde t_N^2} \right] 
 	\approx \frac{1}{ N^2 \, \mathrm{E}[\Delta t]^2} \quad\qquad 
 	\mathrm{Var}\left[\frac{1}{\Delta \tilde t_N} \right] 
 	\approx \frac{ \mathrm{Var}[\Delta t]}{ N^3 \,\mathrm{E}[\Delta t]^4}
 \end{equation}
Given the first of the above relations, the expectation value of $\mathcal{A}_N$ in \eqref{eq:mean_AN} can be approximated as 
 \begin{equation}
 	\mathrm{E}[\mathcal{A}_N] \approx \frac{\mathrm{E}[\Delta \Omega]}{\mathrm{E}[\Delta t]}\, ,
 \end{equation}
 while the variance of $\mathcal A_N$ in \eqref{eq:var_AN} is given by 
 \begin{equation}
 	\mathrm{Var} [\mathcal{A}_N] \approx N^{-1} \left[ \frac{\mathrm{Var}[\Delta \Omega]}{\mathrm{E}(\Delta t)^2} + \frac{\mathrm{E}[\Delta \Omega]^2}{\mathrm{E}[\Delta t]^4} \mathrm{Var}[\Delta t] \right]\, .
 \end{equation}

%%%%%%%%%%%%%%%%%%%%%%%%%%%%%%%%%%%%%%%%%%%%%%%%%%%%%%%%%%%
% Derivation of the moment of inertia constraint
%%%%%%%%%%%%%%%%%%%%%%%%%%%%%%%%%%%%%%%%%%%%%%%%%%%%%%%%%%%

\section{Derivation of the Moment of Inertia Constraint}
\label{app:moi_constraint}

Let us write the dynamics of the total angular momentum~$L$ of the pulsar (neglecting temporal variations in the total moment of inertia $I$) as  
\begin{equation}
\partial_t \, L[\Omega_p, \Omega_{np}] = \dot{\Delta L} + I \dot\Omega_p = -I |\dot{\Omega}_{\infty}|\, ,
\label{eq:momentum_conservation}
\end{equation}
where we assumed that the normal component is rigidly rotating with angular velocity $\Omega_p$, while $\Delta L$ is the angular momentum reservoir due to the (non-rigid) angular velocity lag $\Omega_{np}$ between the superfluid and the normal component \cite{antonelli+2018}. Clearly, \eqref{eq:momentum_conservation} is valid only if we are assuming that the rigid component and the fluid one share a common rotation axis.

It is useful to formally divide $\Omega_p$ and $\Delta L$ into the contributions due to the smooth relaxation ($R$) and an impulsive one from glitches ($G$), 
\begin{equation}
\dot\Omega_p = \dot\Omega_p^G + \dot\Omega_p^R\, \qquad 
\dot{\Delta L} = \dot{\Delta L}^G + \dot{\Delta L}^R \, .
\end{equation}
During glitches, we have $ \dot{\Delta L}^G <0$ and $ \dot{\Omega}_p^G >0$, while for the rest of the time, $ \dot{\Delta L}^R >0$ and $ \dot{\Omega}_p^R <0$. 
We can average \eqref{eq:momentum_conservation} over a long time interval $T_\textrm{obs}$ to get 
\begin{equation}
\langle \dot{\Delta L}^G \rangle + I \langle \dot \Omega_{p}^G \rangle +
\langle \dot{\Delta L}^R \rangle + I \langle \dot \Omega_{p}^R \rangle 
= - I |\dot\Omega_\infty|\, .
\label{eq:averaged}
\end{equation}
We can simplify the equation above by making two observations.
Firstly, due to the angular momentum conservation during a glitch, we must have that (note that $\langle \dot\Omega_p^G \rangle=\mathcal{A}$ by definition) 
\begin{equation}
\langle \dot{\Delta L}^G \rangle + I \langle \dot \Omega_{p}^G \rangle = 0 
\qquad \Rightarrow \qquad
\langle \dot{\Delta L}^G \rangle = -I \mathcal{A}\, .
\label{eq:act_cond_1}
\end{equation}
Secondly, over long time scales, the star spins down as a whole: the reservoir $\Delta L$ fluctuates, but remains bounded (i.e., $\langle \dot{\Delta L} \rangle = [\Delta L(T_\textrm{obs})-\Delta L(0)]/T_\textrm{obs} \rightarrow 0$ for~$T_\textrm{obs} \rightarrow \infty$), so that  
\begin{equation}
\langle \dot{\Delta L} \rangle =
\langle \dot{\Delta L}^G \rangle + \langle \dot{\Delta L}^R \rangle
\approx 0 
\label{eq:act_cond_2}
\end{equation}
if $T_\textrm{obs}$ is long enough.
Now, we do not know the details of the inter-glitch dynamics $(R)$, but it is possible to set an upper bound to $\dot{\Delta L}^R$ by considering the hypothetical perfect pinning limit $(P)$ in which the vortex creep is completely suppressed: $0<\dot{\Delta L}^R < \dot{\Delta L}^P$ and $\dot\Omega_p^P<\dot\Omega_p^R<0$. In this way, we obtain the constraint  
\begin{equation}
 \langle \dot{\Delta L}^P \, \rangle\, > \, I \mathcal{A} 
 \qquad \text{or} \qquad 
 \langle \dot{\Omega}^P_p \, \rangle + \mathcal{A} < - |\dot{\Omega}_\infty|\, .
 \label{cristobaldo}
\end{equation}
Note that it is not important whether or not the perfect pinning is realized in a real pulsar: what we are interested in is giving an estimate of $\dot{\Delta L}^P$, or $\dot{\Omega}^P_p$, in order to set a limit on the real averaged dynamics in \eqref{eq:averaged}. 

To do this, we follow the analysis in \cite{antonelli+2018} and assume that the superfluid component is non-rigidly rotating, so that the angular velocity lag is\footnote{We define as $(r, \vartheta, \varphi)$ the spherical coordinates, with $\vartheta$ and $\varphi$ being the polar and azimuthal angles, respectively; the cylindrical coordinates are defined as $(x,z,\varphi)$, with $z$ being the $\vartheta=0$ axis.%please check the use of footnotes
}
$\Omega_{np}(x,z,t)$. 
Since we are considering a spacetime with circular symmetry, the spacetime metric reads 
\begin{equation}
	\mathrm{d} s^2 = - e^{2 \Phi} \mathrm{d}t^2 + e^{2 \Lambda} \mathrm{d} r^2 + r^2 \mathrm{d} \vartheta^2 + r^2 \sin^2\vartheta \, \left(\mathrm{d} \varphi - \omega\mathrm{d}t \right)^2 \,.
\label{eq:hartle_metric}
\end{equation}
Following the analysis in \cite{antonelli+2018}, the angular momentum reservoir $\Delta L$ is  
\begin{equation}
	\Delta L[\Omega_{np}] = 
	 \int d^3x \, x^2 \, e^{\Lambda-\Phi} \left(\rho\, + P \right) y_n \, \Omega_{np} \, ,
	\label{eq:DL}
\end{equation}
while we define the total moment of inertia as 
\begin{equation}
	I = \int d^3x \, x^2 \, e^{\Lambda-\Phi} \left(\rho+ P\right) (1-\tilde{\omega})\, ,
\label{eq:I_rot}
\end{equation}
where $\tilde{\omega} = \omega/\Omega_p$, $P$ is the pressure, $\rho$ is the internal energy and $y_n$ is the fraction of neutrons in the region where pinning is possible.
However, in \eqref{eq:momentum_conservation}, we explicitly neglected the temporal variations of $I$, but this is in sharp contrast with the fact that $\tilde{\omega} = \tilde{\omega}(r,\theta,\Omega_p,\Omega_{np})$; see \eqref{eq:I_rot}. 
Therefore, in the following, we will ignore the time variations of the rescaled frame drag $\tilde\omega$ for simplicity, although they may play a relevant role \cite{andersson_comer2001,sourie+2017}. 
This amounts to neglecting the dependence of $\tilde{\omega}$ on the small lag $\Omega_{np}$. In this way, in the limit of slow rotation, we have that $\omega(r,\Omega_p)=\tilde{\omega}(r)\Omega_p$, where $\tilde{\omega}(r)$ is a fixed radial function \cite{hartle1967}. For our numerical estimates, we calculate $\tilde\omega(r)$ by following the slow rotation prescription of \citet{hartle1967} (in particular, all the structural functions $\rho$, $P$, $\Lambda$, $\Phi$, $\varepsilon_n$ and $y_n$ are radial functions that can be obtained by solving the TOV%define if appropriate
 equations \cite{antonelli+2018,gavassino+2020_mf}).

Since we have to invoke the perfect pinning condition $(P)$, it is convenient to proceed by using the lag between the normal component and the superfluid momentum, defined as \cite{antonelli+2018,gavassino+2020_mf,montoli_magistrelli+2020}  
\begin{equation}
	\Omega_{vp}=(1-\varepsilon_n ) \Omega_{np} 
	\qquad \Rightarrow \qquad
	\Delta L[\Omega_{np}] = \Delta L[\Omega_{vp}/(1-\varepsilon_n )] \, , 
	\label{eq:omegavp}
\end{equation}
where $\varepsilon_n$ is the entrainment parameter \cite{chamel2013,chamel2017entrainment}.
All the relations obtained until now are valid also in the presence of entrainment;
the only difference is that $\Delta L$ is written as a function of the rescaled lag $\Omega_{vp}$ instead of $\Omega_{np}$, by using Equation~\eqref{eq:omegavp}.
Let us also define the lag derivative $\partial_t \Omega_{vp}^P>0$, which sets an upper limit on the value of $\partial_t \Omega_{vp}^R = (1-\varepsilon_n )\partial_t \Omega_{np}^R$ and is realised when the vortex configuration is perfectly pinned. 
The time derivative of $\Omega_{vp}^P$ reads \cite{antonelli+2018} 
\begingroup\makeatletter\def\f@size{8.5}\check@mathfonts
\def\maketag@@@#1{\hbox{\m@th\normalsize\normalfont#1}}%
\begin{equation}
 \partial_t ( \Omega_{vp}^P + \Omega_{p}^P -\omega(\Omega_{vp}^P , \Omega_{p}^P) ) = 0
 \quad \Rightarrow \quad
	\partial_t \Omega_{vp}^P = - \dot \Omega_p^P + \partial_t \omega(\Omega_{vp}^P , \Omega_{p}^P) \, 
	\approx - \dot \Omega_p^P (1 - \tilde{\omega}) \, .
	\label{eq:dotOmegaVpHartle}
\end{equation}
\endgroup
Clearly, Equation \eqref{eq:momentum_conservation} must hold also if the perfect pinning condition is assumed in the inter-glitch time, namely 
\begin{equation}
	\langle \Delta L [ \partial_t \Omega_{np}^P ] \rangle + I \dot \Omega_p^P = - I | \dot \Omega_\infty |\, .
\end{equation}
Employing Equation~\eqref{eq:dotOmegaVpHartle} in the above relation, we find  
\begin{equation}
	\dot \Omega_p^P ( I - I_v) = \langle \dot \Omega_p^P \rangle ( I - I_v) 
	= - I | \dot \Omega_\infty |\, ,
	\label{eq:dotOmegapP}
\end{equation}
where $I_v = \Delta L [ 1-\tilde{\omega}] $, which coincides with the formula given in \eqref{eq:Iv}.
Finally, let us now come back to Equation~\eqref{cristobaldo}: by using the above result for $\langle \dot \Omega_p^P \rangle$, we finally obtain 
\begin{equation}
 \mathcal{A} - \frac{I}{I-I_v} | \dot \Omega_\infty| < - | \dot \Omega_\infty| \, ,
\end{equation}
which is equivalent to the constraint in Equation~\eqref{eq:act_constraint}.
Let us remark that the present result is based on the quasi-stationary approach used in \cite{antonelli+2018,gavassino+2020_mf}, an approximation that can be justified by the fact that the glitch rise time is expected to be orders of magnitude larger than the hydrodynamical time scale, as discussed in \cite{sourie+2017}.
Of course, an analogous result can be obtained in a completely rigorous way (since there is no need to find approximations for the frame drag $\omega(\Omega_p,\Omega_{np})$ and its temporal derivative) also in a Newtonian context: the form is still the one in \eqref{eq:act_constraint}, and $I_v$ is given by \eqref{eq:Iv}, but with $\omega=\Lambda=\Phi=0$.

%%%%%%%%%%%%%%%%%%%%%%%%%%%%%%%%%%%%%%%%%%
\reftitle{References}


\externalbibliography{yes}
\bibliography{biblio}

%% for journal Sci
%\reviewreports{\\
%Reviewer 1 comments and authors’ response\\
%Reviewer 2 comments and authors’ response\\
%Reviewer 3 comments and authors’ response
%}

%%%%%%%%%%%%%%%%%%%%%%%%%%%%%%%%%%%%%%%%%%
\end{document}